\documentclass[a4paper,fleqn]{cas-sc}

\tnotemark[1]
\tnotetext[1]{This is a preprint.}
\renewcommand{\printFirstPageNotes}{}

\usepackage[numbers]{natbib}

\usepackage{cleveref}

\usepackage{geometry}
\usepackage[
	textsize=scriptsize,
	colorinlistoftodos,
]{todonotes}

\reversemarginpar
\usepackage{tabularray}

\usepackage{tikz}
\usetikzlibrary{shapes.geometric, arrows, arrows.meta, positioning}

\tikzstyle{block} = [rectangle, draw, fill=blue!20, text centered, rounded corners, minimum height=3em, minimum width=6em]
\tikzstyle{line} = [draw, -latex']

\tikzstyle{startstop} = [rectangle, rounded corners, minimum width=3cm, minimum height=1cm,text centered, draw=black, fill=red!30]
\tikzstyle{process} = [rectangle, minimum width=3cm, minimum height=1cm, text centered, draw=black, fill=orange!30]
\tikzstyle{decision} = [diamond, minimum width=3cm, minimum height=1cm, text centered, draw=black, fill=green!30]
\tikzstyle{arrow} = [thick,->,>=stealth]

\def\tsc#1{\csdef{#1}{\textsc{\lowercase{#1}}\xspace}}
\tsc{WGM}
\tsc{QE}
\begin{document}
\let\WriteBookmarks\relax
\def\floatpagepagefraction{1}
\def\textpagefraction{.001}

\shorttitle{Adaptive Learning Pipeline for Comprehensive AI Analysis}    

\shortauthors{Torka et~al.}  

\title [mode = title]{Adaptive Learning Pipeline for Comprehensive AI Analysis}  

\tnotemark[1] 

\tnotetext[1]{BMBF, BSI funded} 


\author[1]{Simon Torka}

\cormark[1]

%\fnmark[<footnote mark no>]

\ead{simon.torka@dai-labor.de}

\ead[url]{https://dai-labor.de/}

%\credit{Writer}

\affiliation[1]{organization={DAI-Labor, Technische Universität Berlin},
            addressline={Ernst-Reuter-Platz 7}, 
            city={Berlin},
%          citysep={}, % Uncomment if no comma needed between city and postcode
            postcode={10587}, 
            country={Germany}}

\author[1]{Sahin Albayrak}

% Footnote of the second author

% Email id of the second author
\ead{sahin.albayrak@dai-labor.de}

\ead[url]{https://www.dai-labor.de}

%\credit{Inspiration}

\cortext[1]{Corresponding author}

%\fntext[1]{ }


\begin{abstract}
\ignorespaces
The advancement of artificial intelligence (AI) technologies has significantly increased the complexity of AI pipelines, involving intricate algorithms as well as stages such as data collection, preprocessing, training, evaluation, and visualization. To ensure effective and accessible AI solutions, it is crucial to design pipelines accommodating diverse user groups, including experts, specialists from various fields, and novices. The usability and functional scope vary based on user groups, and user experience is pivotal for building trust in AI systems. This paper emphasizes the need for adaptive AI pipelines to meet growing complexity and diverse user demands and offers a solution. Introducing ALPACA as a comprehensive AI pipeline highlights its potential to enhance usability, trust, and security in AI systems. ALPACA and similar systems facilitate the integration of AI into daily life, enabling diverse user groups to harness AI capabilities while mitigating risks and challenges.

In this paper, we present ALPACA (Adaptive Learning Pipeline for Advanced Comprehensive AI Analysis), a comprehensive and holistic AI pipeline framework. We showcase ALPACA's capabilities through an Android similarity detection application. ALPACA accommodates diverse user groups by integrating techniques such as linking visual and code-based development and making important phases such as data collection, processing, training, evaluation and visualisation easily accessible.

The investigation will comprehensively examine the design principles and frameworks underlying ALPACA. The system employs Celery with a Redis backend for efficient task management and scalability. MongoDB handles data storage across pipeline stages seamlessly. The system is hosted on a Kubernetes server for cloud-based advantages such as scalability, availability, and resource efficiency. Aside from that, future versions will be able to integrate modern techniques like federated and continuous learning, along with explainable AI methods.

\end{abstract}


% \item

% Keywords

\maketitle

% Main text

\section{Introduction}
\label{sec:intro}

In the rapidly evolving landscape of artificial intelligence (AI), the ability to comprehensively analyze and understand complex data is paramount. However, the ever-increasing complexity of real data requires sophisticated AI pipelines that seamlessly integrate different stages such as data collection, preparation, model generation, and evaluation. Such a pipeline can be illustrated as a chain of distinct yet interdependent stages, each contributing to the overarching goal of turning data into actionable intelligence. Like a well-coordinated symphony, these stages require harmonious collaboration to achieve optimal results. The concept of AI pipelines therefore represents more than just a linear progression; it signifies the orchestration of diverse processes to accomplish a larger purpose. At the core of this revolution lie AI pipelines, intricate networks of interconnected data processing and analysis steps designed to transform raw data into meaningful insights or outcomes using AI techniques. The evolution of simple AI models to adaptive, systematic AI pipelines has ushered in a new era of data-driven decision-making by solving complex tasks in an ever-changing environment. Making AI understandable, accessible and usable by everyone in every domain requires a domain-independent, easy-to-use pipeline architecture that can be integrated into a complex ecosystem of experts and non-experts. However, the design and implementation of such pipelines often prove challenging due to the intricate interplay of technical components and the diverse requirements of different application domains.

The main objective of this document is to provide a comprehensive description of the ALPACA (Adaptive Learning Pipeline for Advanced Comprehensive AI Analysis) system and to explain the underlying architectural principles as well as the technical details of the operational phases. Through a detailed technical investigation of each pipeline phase and a holistic integration of supporting LLM techniques, ALPACA aims to unify and improve the accessibility of AI for different user groups. In addition, the system should be able to autonomously detect the available hardware and make optimal use of the available resources. This should allow even inexperienced users to benefit from hardware-related improvements such as optimal use of available GPUs to increase computing power. This also improves scalability across different systems and helps reduce costs, thereby reducing the initial burden of hardware investments and promoting the widespread adoption of this technology. In addition, the system should also be able to distinguish a production system from a training system and safely scale solutions developed in the training system to a production system.

The system is designed to support different user groups with different expertise and expectations through a web-based user interface, thus fostering a more comprehensive AI ecosystem. It takes into account the different knowledge levels of users and facilitates the complex orchestration of configurable AI pipeline stages. This user-centric approach can improve the adoption of AI in civil society by making it widely accessible. The following user group is addressed:

\begin{itemize}

\item \textbf{Data scientists and AI experts:} This group, which has a solid foundation in AI and machine learning, benefits from ALPACA through prompt-based optimization of models and workflows. In addition, added value can be gained from the prompt-based provision and administration of a controllable and secure environment for testing various models and hyperparameters during the development phase. Another advantage is the ability to analyze the usable hardware and to prompt-based rollout of finished AI systems. In addition, knowledge exchange and collaborative work with other trades can be benefited.

\item \textbf{Specialists in other areas:} Subject-specific experts with limited AI knowledge can use LLM-supported AI pipelines to bridge the gap between their expertise and advanced AI techniques. These systems democratize AI by providing user-friendly interfaces and automated workflows, thus facilitating the integration of subject-specific knowledge with AI skills.

\item \textbf{Students and laypeople:} Even people without specific AI knowledge can benefit from such systems. By incorporating interactive tutorials, example projects and guided exercises, these systems can also be seen as educational platforms that allow users to explore AI through hands-on learning and real-world challenges. This inclusive approach not only promotes learning but also contributes to a wider adoption of AI within this user group.

\end{itemize}

In order to test the applicability, application scenarios are defined, based on which the performance of ALPACA is tested in thought experiments. These use cases include research projects we have carried out in the past in the areas of third-party research and industrial research. It should be noted that ALPACA serves as a generic AI pipeline that can be used in various areas, but is mostly confronted with tasks in the area of IT security. The results of these experiments lay the foundation for the development of ALPACA. ALPACA is intended to help answer the following research questions:

\begin{itemize}

\item How can LLM techniques be used effectively in AI pipelines?

\item Is it possible to develop a universally applicable AI pipeline that remains usable for different user groups?

\item Can an AI ecosystem be enriched by incorporating inputs from different user groups?

\item How can such a pipeline be optimized for experts and yet remain usable for laypeople?

\item Can the output and decision-making processes of LLM-based AI pipelines be made comprehensible for all user groups?

\item Can a system like ALPACA lead to an ecosystem in which different user groups can participate and different neglected domains can also be penetrated by AI?

\end{itemize}

The paper is structured as follows. Chapter 2 provides an overview of the state of the art and summarises the knowledge required for the work. Chapter 3 describes the requirements specification of ALPACA, while Chapter 4 explains the ALPACA system. Chapters 5, 6 and 7 present the data acquisition pipeline (stage 1), the data pre-processing pipeline (stage 2) and the AI model pipeline (stage 3). Chapter 8 deals with legal and ethical considerations. Chapter 9 briefly summarises the findings of the paper. Chapter 10 outlines the future potential and extensibility of the system.

\section{State of the Art}
\label{sec:stateoftheart}

Artificial intelligence (AI) pipelines play a critical role in the modern landscape of AI systems by facilitating their efficient development, deployment, and maintenance. These pipelines orchestrate a series of sequential steps that transform raw data into valuable output \cite{Baylor.2017}. This output ranges from refined datasets and trained AI models to complete predictive systems and includes stages such as data preprocessing, model training, evaluation, data prediction, and deployment to cloud or edge infrastructure \cite{Steidl.2023}. A key benefit of AI pipelines is their ability to automate repetitive and time-intensive tasks inherent in AI development, including data refinement, feature engineering, and model selection \cite{Radanliev.2023}. By mechanizing these processes, AI pipelines significantly reduce the time and cognitive demands and simplify the use of AI to solve real-world problems \cite{Radanliev.2023}. According to \citeauthor{Hummer.2019}, this time-saving advantage can be further extended by using parallel computations, which can save up to 7/% of the time.

Building AI pipelines can involve various levels of sophistication, from single, highly specialized pipelines designed for specific use cases to multi-functional, cloud-based AI pipelines. In recent years, the field of open source ML pipelines has seen significant growth, with tools such as FluxCD \cite{FluxCD.20231210} specializing in continuous delivery for seamless updates of containerized applications in Kubernetes clusters. Various tools and frameworks such as DVC (Data Version Control) \cite{DVC.20231210b}, Git Large File Storage \cite{GitLargeFileStorage.20231025}, Apache Hadoop \cite{ApacheHadoop.20231209}, Apache Airflow \cite{ApacheAirflow.20231206}, Apache Spark \cite{ApacheSpark.20230922}, Apache Flink \cite{ApacheFlink.20231202}, Apache NiFi \cite{ApacheNiFi.20231128}, Prefect \cite{Prefect.20231210} and Luigi \cite{Luigi.20210129} together provide a versatile toolkit that covers a wide range of use cases and takes into account different preferences and requirements.

Additionally, the field has seen significant growth in generic open-source ML Pipelines, exemplified by MLFlow \cite{MLflow.20231207}, a versatile platform for end-to-end machine learning and lifecycle management; DataRobot \cite{DataRobotAIPlatform.20231207}, an automated machine learning platform streamlining model development; H2O.ai \cite{H2O.ai.20231210}, an open-source framework emphasizing user-friendly model building and scalability; andAutoAI \cite{Radanliev.2023}, an advanced autonomous AI design based on cutting-edge algorithms.

Specialized pipelines have emerged from recent research, such as KoopaML \cite{GarciaHolgado.2022} and IPMP \cite{Frey.2020} in the medicine and health care sector, as well as proposals from companies like Uber's Michelangelo, addressing the complete lifecycle of AI models, and Facebook's FBLearner, focusing on internal ML initiatives.

Cloud-based pipelines, driven by global players like Google Cloud Vertex AI Pipelines \cite{google.vertex.2023}, Amazon SageMaker \cite{AmazonWebServicesInc.26.10.2023}, Microsoft Azure Machine Learning \cite{Microsoft.Azure.31.10.2023b}, IBM Watson Studio \cite{IBM.Watson.2022b}, Databricks \cite{Databricks.31.10.2023b}, Kubeflow \cite{Kubeflow.20231210b} and Google's TFX \cite{Baylor.2017}, offer comprehensive platforms for establishing, deploying, and overseeing end-to-end machine learning workflows. However, the high cost and complexity of these pipelines present barriers to entry for many users. 

In contrast, AI4EU \cite{AIBuilder.20231123, AIBuilderDocumentation.20231219, AIWatch.31.10.2023, AI4EU_Resultsplatform.20231114, AI4EU_Vision.2023}, a European initiative, was developed to transfer knowledge from research to business applications. It promotes a “European on-demand platform and ecosystem for artificial intelligence” \cite{AIWatch.31.10.2023} that “brings the AI community together” \cite{AIWatch.31.10.2023}. In this context, it can be used by experts developing AI components and making them available to the community, as well as by laypeople without AI knowledge.

Despite these advances, none of the pipelines developed has provided a universal solution applicable in all contexts. In addition, no AI pipeline is known to date that addresses AI experts, domain experts and laypeople equally by integrating LLM techniques and offers comprehensive support over the entire product lifecycle of an AI. Consequently, research on AI pipeline design remains an active area of investigation, aiming to enhance both efficiency and efficacy \cite{Steidl.2023}. To close this gap \citeauthor{Steidl.2023} identifies key challenges in systematizing, unifying, and expanding AI pipelines, emphasizing the need for the identification and structural organization of central conceptual ideas. They note a lack of end-to-end pipelines covering the complete AI lifecycle from data collection to AI model deployment and monitoring in production \cite{Steidl.2023}. The study uncovers deficiencies in all phases of AI pipelines, from data collection and versioning to operating the entire pipeline in a business context. To address these issues, the authors propose a taxonomy of a four-stage AI pipeline, encompassing data handling, model learning, software development, and system operations. We will further extend this approach by integrating LLM.

\section{ALPACA Requirements Specification}
\label{sec:keyfeatures}

To achieve the goals mentioned above, ALPACA must fulfill several important functions. 
One focus is on the \textbf{user interface}, which must be as easy to use as possible. This also means that all visualizations of the pipeline stages, namely data acquisition, data preprocessing, model training, model evaluation, and data prediction, must be easily accessible, understandable, and configurable.

Another key feature regarding the user interface is \textbf{handling different user levels}. The system must acknowledge different user levels, catering to novice users exploring AI and advanced developers extending the system's capabilities. This user-centric design enhances usability and accessibility.

To ensure future viability, another key feature is the \textbf{maintainability and expandability} of the entire system. It must be ensured that developers can effortlessly integrate new functions into the system without adapting existing parts. The GUI and the underlying architectural principle must be based on mechanisms and strategies that enable quick and easy integration of new functions. For example, the user interface must offer mechanics with which new functions can be seamlessly integrated without having to change the GUI. Reflections,  a mechanism allowing dynamic examination and modification of program structure at runtime, can be used so that users can easily benefit from new pre-processing steps and AI algorithms. For this purpose, the system must offer base classes and interfaces that are based on reflections, which are optimized for automatic GUI generation. This mechanism eliminates the need for manual GUI development and simplifies the customization of preprocessing pipelines. Additionally, the provided base classes request developers to provide a useful description of each feature and each parameter implemented. This can assist inexperienced users in using the functions provided.

The ALPACA system's \textbf{scalability} plays a crucial role in achieving its adaptability, usability, and robustness because it enables the system to adapt to evolving requirements and accommodate larger datasets. This should be ensured by deploying the system on a Kubernetes server, which also enables the efficient allocation of resources and thus meets growing computing requirements. For example, all pods can be orchestrated by Kubernetes, allowing it to scale horizontally by deploying multiple containers across clusters. This elasticity ensures optimal resource utilization and performance even as demands fluctuate.

But all of these features are useless if the results are not \textbf{reproducible and trustworthy}. To ensure this, the system must capture and store all artefacts necessary for reproducibility. On the one hand, this includes all user interactions, but on the other hand, it also includes all generated data. While the user interactions can be saved in XML data, the artefacts must be stored in a compressed manner that is as memory-saving as possible. Furthermore, it must be ensured that the log data is passed on across all pipeline stages and enriched with the user interactions of each pipeline stage. This, on the one hand, facilitates the seamless scaling of the prediction component; on the other hand, it enables reproducibility across all stages. The resulting close documentation of all data also enables data to be replicated and validated. Guaranteeing the essential cornerstones of scientific work strengthens trust in the system's analyses and findings. 

In the following sections, we begin a detailed examination of our proposed AI pipeline, ALPACA, and highlight its transformative potential. 
\section{The ALPACA System}
\label{sec:system_architecture}

\begin{figure}[h]
  \centering
  \begin{minipage}{0.49\linewidth}
    \centering
    \includegraphics[width=\linewidth]{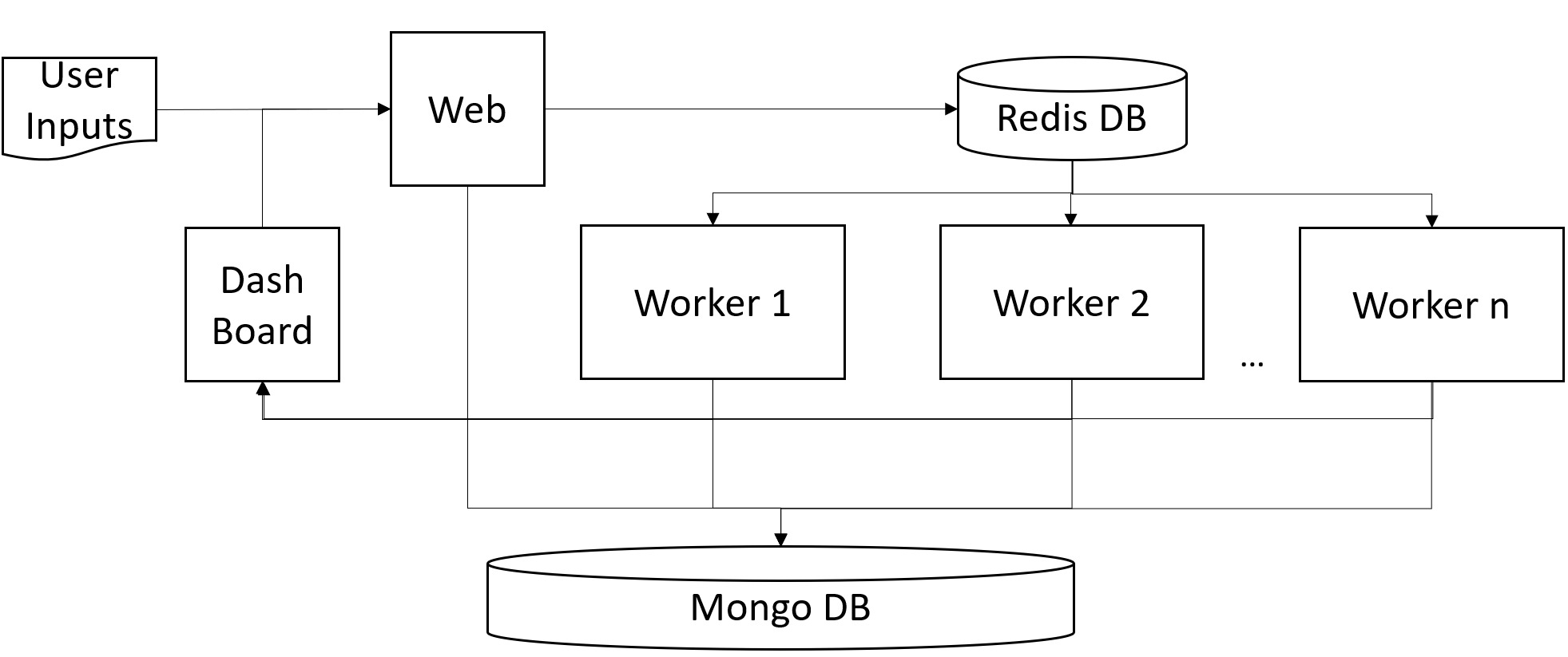}
    \caption{Alpaca's Dataflow.}
    \label{fig:Architecture}
  \end{minipage}
  \hfill
  \begin{minipage}{0.49\linewidth}
    \centering
    \includegraphics[width=\linewidth]{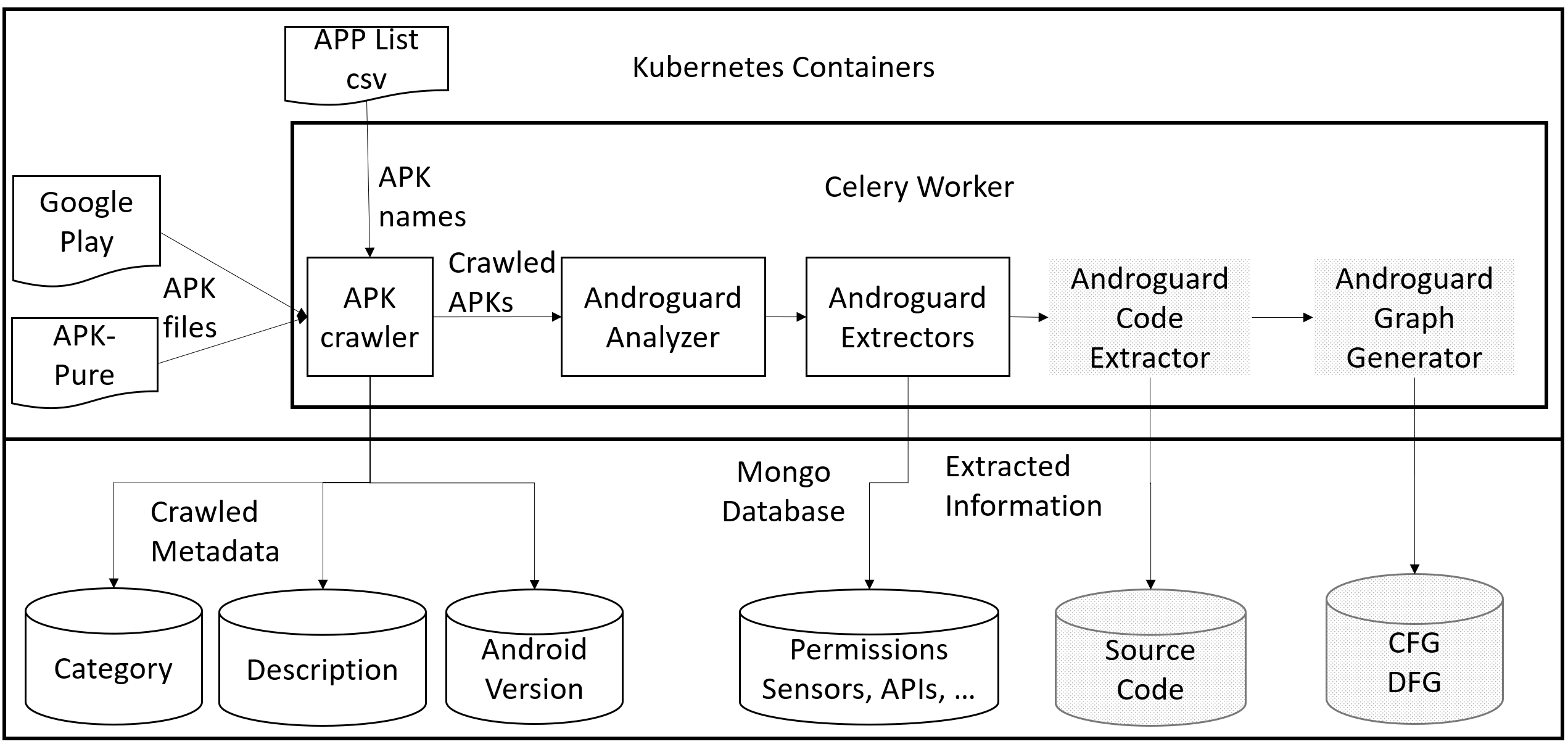}
    \caption{Alpaca's Container Architecture.}
    \label{fig:Container}
  \end{minipage}
\end{figure}

The system is designed as a versatile and modular platform to address the complexity of AI analysis and increase accessibility to this new disruptive technology. To take into account the complexity of data analysis pipelines and enable seamless integration of the data collection, pre-processing, and AI model training phases, much emphasis was placed on a modular, cloud-based system architecture. This enables comprehensibility and access to AI for different user groups. Great importance was placed on ease of use and expandability. To shed light on these facts, we delve deep into the system architecture and examine individual pipeline stages, technical components and complex functionality. At its core, ALPACA is based on the principle of adaptability. To comprehend the system's architecture, we initiate our exploration from a user's perspective. The complete workflow is governable and configurable through a web UI. Initiated by users, tasks are orchestrated by Celery and executed through interacting Kubernetes pods, thereby ensuring the automated processing of individual AI analysis steps. The data flow, initiated by user input and traversing through processing workers to data storage in the database, is depicted in Figure \ref{fig:Architecture}. Additionally, the specific pod architecture for our use case is presented in Figure \ref{fig:Container}. The modular structure allows developers to quickly and easily expand the system and adapt it to new scenarios. This means that ALPACA can also be used as the basis for an AI ecosystem, similar to the AI4EU project \cite{AI4EU_Vision.2023, AIWatch.31.10.2023, AI4EU_Resultsplatform.20231114}. Our research shows how easy-to-use AI pipelines can help facilitate and democratize access to AI capabilities. This transformative approach could reshape the landscape of AI analytics pipelines and stimulate a new wave of research, development, and use in artificial intelligence.

The ALPACA system is designed to integrate both existing and new components into the system seamlessly. The focus is on modularity and expandability, as well as the user-friendly operation of the system. The system's architectural design embodies these principles, fostering a cohesive amalgamation of its diverse components. 

The pipeline consists of the three basic phases of data acquisition, data preparation, and model training. The first stage involves the implementation of the data collection pipeline, a critical stage that exhibits a pronounced correlation with the underlying problem to be addressed. The specific use case under consideration in this research pertains to the identification of functionally similar Android applications. Within this context, the primary objective of the first stage of the pipeline is to acquire data from APK repositories, employing sophisticated web scraping techniques. Furthermore, this phase of the pipeline is tasked with the extraction of pertinent features from this fundamental APK data. To achieve this feature extraction, the AndroGuard tool \cite{androguard.readthedocs.io.29.01.2021} is employed as a means of facilitating the extraction of these discriminative features. 
The second stage is dedicated to data preprocessing, which has the ability to move away from the specific use case and enable a generic perspective. In it, the collected features will be subjected to careful refinement to convert them into a format suitable for training AI models. This stage includes selecting, merging, and transforming the data. 
The third stage assumes responsibility for the artificial intelligence (AI) model and, similar to the antecedent second stage, can be regarded with relative independence from the specific scenario under consideration.

This chapter continues to provide a comprehensive overview of ALPACA's front-end and back-end architecture and highlights the seamless integration of various technologies. It becomes clear that ALPACA is more than just an analysis tool; it is evidence of the advances in adaptive AI pipelines. Through its adaptability, ALPACA not only addresses the existing limitations of modern AI pipelines as outlined in \cite{Steidl.2023}, but also pioneers novel solutions to the challenges facing the AI community. ALPACA's architecture relies on several software components that form the system's backbone. These are presented in the following subchapters.

\subsection{Technical Fundament}
%%Python3 \cite{Python.org.08.11.2023} and its extensive library landscape form the basic framework of the system. As a robust container orchestration platform, Kubernetes \cite{Kubernetes.08.11.2023} handles the efficient provisioning and management of system resources.

Python3 \cite{Python.org.08.11.2023} and its rich library ecosystem constitute the foundational framework for ALPACA. It was chosen for its widespread adoption in the AI community, which facilitates seamless integration with popular ML libraries. As a container orchestration platform, Kubernetes \cite{Kubernetes.08.11.2023} was selected for its industry-wide adoption and proven efficiency in dynamically provisioning and managing system resources for scalable AI workflows.

\subsection{Backend Technologies}
The foundation of any AI pipeline is data management and storage. At the software level, Pandas \cite{McKinney.2010} data frames are used, increasing the efficiency and effectiveness of data processing. MongoDB \cite{MongoDB.08.11.2023} serves as the system's database, providing robust and flexible storage for collected app data, preprocessed datasets, trained models, and user interactions.
Task coordination for container-based workers is handled by Celery \cite{celery.03.09.2023}, a distributed task queue for efficient task management that ensures streamlined and seamless handling of data collection, preprocessing, and model training. Redis \cite{Redis.08.11.2023} was chosen as the Celery backend due to its ability to work in RAM, which significantly increases the performance of the system. This orchestration not only facilitates smooth operations but also contributes to scalability, reproducibility of results, and ease of use.

To practically demonstrate ALPACA's capabilities, an app similarity detection system was implemented. A Selemium-based \cite{selenium.30.09.2023} crawler is used in conjunction with the  Firefox Geckodriver in headless mode \cite{GitHub.08.11.2023} to collect app information. The collected APKs are taken over by Androguard, which then takes over the extraction of essential app attributes and functions for analysis. This demo scenario demonstrates ALPACA's adaptability to real-world AI use cases.

On the AI side, all neural networks are implemented using Keras TensorFlow \cite{TensorFlow.02.11.2023}. In addition, Scikit-Learn \cite{scikit-learn.08.11.2023} is used to implement the classic non-neural models. This versatility allows researchers to explore a wide range of AI techniques within the platform.

Together, these frameworks form a cohesive ecosystem that gives developers and users the ability to perform comprehensive AI analysis.

\subsection{Frontend Technologies}
The user-friendly frontend is based on the high-level Python web framework Django \cite{DjangoProject.08.11.2023} and leverages its extensive user interaction features. Users interact with the system through a clear and intuitive website, making it easier to access for different user groups. The powerful, interactive data visualization tool Plotly Dash \cite{dash.plotly.08.11.2023} is used to create interactive data visualizations. It serves as a gateway for user interaction and provides an intuitive interface for different user groups. This allows users to explore the nuances of working with and on AI models through dynamic and engaging graphical representations. Finally, Celery Flower \cite{celery.flower.17.06.2023} enables seamless real-time monitoring of Celery employees, improving the user experience and providing transparency in task processing.

\subsection{Frontend Vizualisations}

\begin{figure}[h]
  \centering
  \begin{minipage}{0.75\linewidth}
    The ALPACA system's frontend, built on Django and Plotly Dash, plays a pivotal role in enabling user interaction with the system's functionality. Key components include:
  \end{minipage}%
  \hspace{0.5cm} % Adjust the space here (1cm in this example)
  \begin{minipage}{0.2\linewidth}
    \centering
    \includegraphics[width=\linewidth]{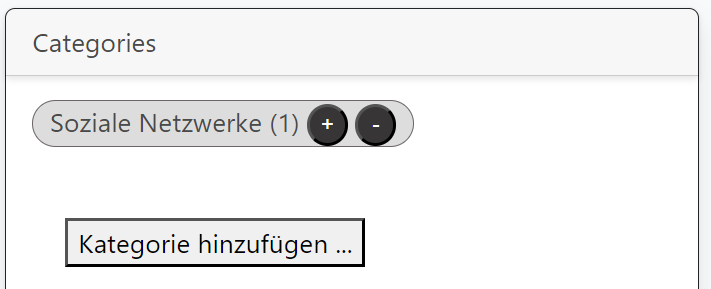}
    \caption{Voting System.}
    \label{fig:Voting_System}
  \end{minipage}
\end{figure}

\begin{itemize}

\item \textbf{APK Upload:} Users can upload single or multiple APKs along with their associated categories. This feature enables crowdsourcing data collection and allows different groups to participate in data collection.

\item \textbf{Celery Dashboard:} A dashboard provides insights into task processing, increasing transparency and monitoring capabilities.

\item \textbf{Database Visualization:} Through visualization, users can explore the data stored in MongoDB. One can see, among other things, all processed APKs, the progress of the feature extraction, as well as the extracted features and the app metadata. In addition, this tab offers an app category voting system (Figure \ref{fig:Voting_System}) and is therefore also part of the crowdsourcing approach.

\item \textbf{Preprocessing and AI:} A user-centred design was developed for the preprocessing and AI stage, which extends across all phases of the pipeline. Users can access and configure these stages via a user-friendly graphical web interface. The configuration of the individual algorithms is designed to be self-explanatory, which further improves accessibility and user-friendliness. For better reproducibility, all user actions and all artefacts generated in the respective pipeline stage are saved.

\end{itemize}

\subsection{Pipeline Stages of ALPACA}
The architecture of the ALPACA system is based on a modular design consisting of three primary pipeline stages: data acquisition, data preprocessing, and AI modelling. These pipelines interact synergistically to process data and generate actionable insights. The basis for this is the effective data management and the intuitively designed user interface, which increases user-friendliness and transparency and thus secures the core tasks of the ALPACA system. Below, we explain the basics of each pipeline stage. A detailed description of the underlying mechanisms and strategies of the pipeline stages, particularly about data management, usability, reproducibility, and transparency, follows in the chapters \ref{sec:data_collection_pipeline},  \ref{sec:data_preprocessing_pipeline}, and \ref{sec:ai_model_pipeline}.

\begin{itemize}

\item \textbf{Stage 1: Data Collection Pipeline:} The data collection pipeline is the system's initial stage, responsible for sourcing and gathering data from app stores. In our demo scenario, we extract app data from APK repositories using web scraping techniques and leverage AndroGuard for comprehensive app feature extraction.

\item \textbf{Stage 2: Data Preprocessing Pipeline:} The data preprocessing pipeline refines the collected data, preparing it for AI model training. For this purpose, the collected data undergoes preprocessing to shape it into suitable formats for AI model training. Data selection, merging, and transformation steps streamline data for subsequent analysis.

\item \textbf{Stage 3: AI Model Pipeline:} This stage encompasses AI model training, evaluation, and prediction. Users select AI algorithms, configure hyperparameters, and choose the dataset for training and analysis.

\end{itemize}

%% Quellen APK und Play Store, Androguard
%% Im Intro auf die Stages einleiten

\section{The Data Collection Pipeline (Stage 1)}
\label{sec:data_collection_pipeline}

The data collection pipeline represents the foundational stage of the ALPACA system and is responsible for acquiring the basic analysis data. As elucidated earlier, the data collection depends largely on the underlying use case. In the demo scenario of detecting similar Android applications, the data subject to analysis is derived from Android packages (APKs). Given the potential divergence in requirements across various use cases, the configuration of this pipeline stage necessitates adaptation contingent upon the specific demands of the given use case and can be easily exchanged and expanded at ALPACA. The following section delves into the intricacies of the data collection pipeline in the context of Android feature extraction, elucidating the roles of the crawler, analyzer, and extractor in gathering metadata, APIs, and other attributes from APKs.

\subsection{Crawler}

The crawler component forms the initial step in the data collection pipeline and is the gateway to accessing information from external sources. This step fundamentally hinges on the particulars of the individual use case and mandates adaptation on a case-by-case basis. It is encapsulated within a Docker container, providing advantages (Figure \ref{fig:Architecture}). Firstly, it facilitates swift and facile interchangeability, allowing for seamless adjustments to suit the specific requirements of the given use case. Secondly, the stage is equipped with the capability to initiate multiple crawlers as celery workers. Specifically, our crawler's role involves extracting data from two primary sources: APK Pure and the Play Store. Crawled data will be stored in MongoDB. In this demo scenario, APK Pure serves as a repository for collecting APK files, while the Play Store is used to gather some metadata related to the APK. This construction was chosen because of the access limitations of the Play Store. The crawler leverages Selenium and a headless Firefox driver to automate the web crawling process, enabling seamless extraction of APK files and metadata.
The metadata acquired from the Play Store includes categories, app descriptions, size, required Android version, developer information, and app ratings. These attributes provide some simple features to the app similarity analysis that are essential for our use case. However, the primary data source is the Android APK extracted by APK Pure. 
This design ensures that ALPACA's data collection pipeline can be adapted to a variety of scenarios, enabling comprehensive analysis and experimentation in different application scenarios.

\subsection{Analyzer}

After gathering the APK files, the analyzer component, which harnesses the capabilities of AndroGuard, comes into play. This component is also encapsulated in a Docker container, which allows us to scale those workers. The analyzer processes the collected APKs and generates AndroGuard sessions that encapsulate the essential information from the APKs. The analyzer component only generates the Androguard sessions and does not extract any features. These sessions are stored in the MongoDB database, which allows us to handle those sessions by multiple workers at the same time without regenerating every time. The sessions form the basis for further analysis, enabling a granular understanding of the apps under examination.\footnote{Regarding federal German law, it is not allowed to decompile software without the explicit acknowledgement of the copyright owner. If you want to use our tool and recreate the demo-implemented use case, please ensure that you have the required permissions. We distance ourselves from any misuse.}

\subsection{Extractor}

Working in tandem with the analyzer, the extractor component leverages AndroGuard to extract a multitude of critical features from the APKs. The extractor is also dockerized and implemented as a celery worker. In addition, the extractor can specifically extract features that are not yet in the database. This allows us to have an APK processed by multiple workers. In addition, aborted extraction processes can be resumed without having to re-extract all features that have already been extracted. For feature extraction, the extractor uses the sessions generated by the analyzer. It can extract the following features:

\begin{itemize}

\item \textbf{APIs:} The extracted APIs include Android APIs, third-party APIs, and internal function names, offering insights into the app's functionality and potential interactions.

\item \textbf{Manifest:} The AndroidManifest.xml is parsed to extract essential information related to the app's configuration and components.

\item \textbf{Strings:} The strings within the APK and manifest are extracted, providing textual insights into the app's purpose and functionality.

\item \textbf{Intents:} The various intents, including services, receivers, and activities, are identified to understand the app's functionalities and communication mechanisms.

\item \textbf{Permissions:} Permissions requested by the App reveal the level of access the app requests from the Android system, shedding light on potential privacy and security implications.

\item \textbf{Features:} The features utilized by the app are identified, offering insights into its capabilities and interactions.

\item \textbf{Sensors:} Information about the sensors used by the app provides insight into the interactions between the app and the environment.

\item \textbf{Code Extractor:} The Code extractor is designed to extract the APK's source code, offering deeper insights into the app's inner workings.\footnote{This system is disabled due to legal concerns.}

\item \textbf{Graph Generator:} This component generates control and dataflow graphs based on the extracted source code, enabling visual representations of the app's logic and interactions.\footnote{This system is disabled due to legal concerns.}

\end{itemize}

The outlined data collection stage is notably contingent upon the characteristics of the specific use case, yet it serves as a template with broader applicability to alternative scenarios. The construct elucidated herein facilitates the effective coordination of components essential for data collection, furnishing subsequent phases of ALPACA with a robust system. This system is instrumental in generating a feature-rich dataset, serving as the foundational substrate for a comprehensive AI analysis.

\section{The Data Preprocessing Pipeline (Stage 2)}
\label{sec:data_preprocessing_pipeline}

The data processing pipeline serves as a central bridge between raw data collection and AI model training and is much more independent of the specific use case than the data collection stage. This section delves into the intricacies of the data processing pipeline, encompassing data selection, merging, and preprocessing stages, each contributing to the refinement and preparation of data for subsequent analyses. The underlying systems are wrapped in celery-based Docker containers, which makes them scalable. All produced data and the system configurations made by the user are stored in MongoDB. Because this stage is only loosely related to the underlying scenario, it can also be transferred to other scenarios. If necessary, however, this can be quickly adapted and expanded.

\subsection{Data Selector}

\begin{figure}[h]
  \centering
  \includegraphics[width=0.55\linewidth]{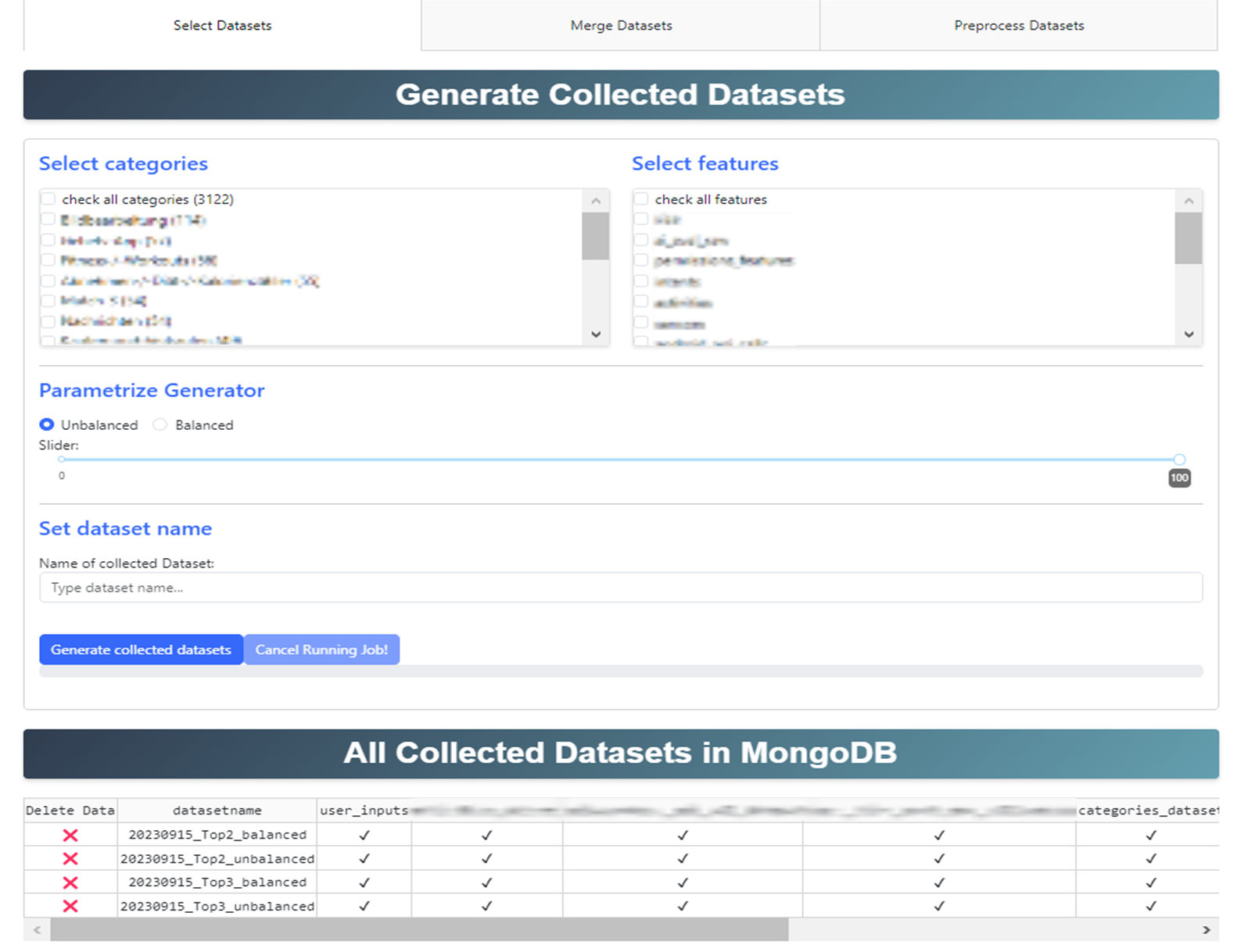}
  \caption{User Interface to Manage the Dataset Selection Process.}
  \label{fig:1.1_Select_Datasets}
\end{figure}

The Data Selector component constitutes the initial step of the Data Preprocessing Pipeline, allowing users to tailor the dataset according to their analysis requirements. In this context, users can configure dataset characteristics, including features, app categories, dataset balance, and inclusion percentages (Figure \ref{fig:1.1_Select_Datasets}).

These selections shape the resulting datasets, which are packaged and saved for future analysis. This component presents users with a set of options to configure the dataset generation process:

\begin{itemize}

\item \textbf{Feature Selection:} Users can choose the specific features to be included in the dataset by selecting from checkboxes corresponding to different attributes.

\item \textbf{Label Selection:} Users can select which labels should be integrated into the dataset, enabling customization based on the focus of the analysis. In the demo case, the labels are represented by app categories.

\item \textbf{Balancing and Subset Selection:} The system accommodates a dynamic approach to selecting a subset of items (0\% to 100\%) based on their occurrence in the dataset. Users can also specify whether the dataset should be balanced (even class distribution) or unbalanced (uneven class distribution).

\end{itemize}

Depending on the configuration, the Data Selector generates individual datasets for each selected feature. All subdatasets are stored in a single compressed file to minimize storage usage. The generated dataset as well as the user input settings are stored in the MongoDB to ensure reproducibility.

\subsection{Data Merger}

\begin{figure}[h]
  \centering
  \includegraphics[width=0.55\linewidth]{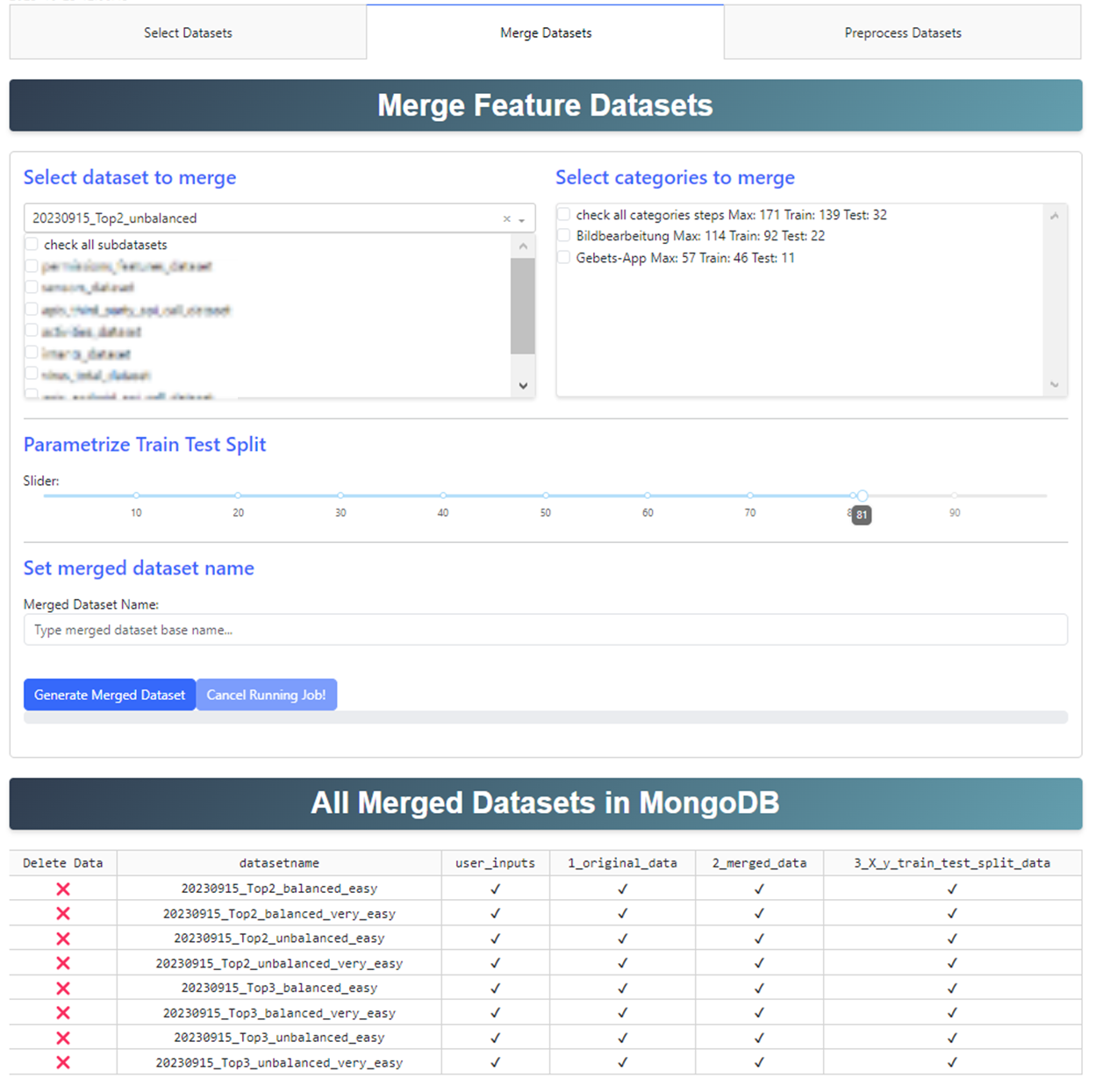}
  \caption{User Interface to Manage the Dataset Merger Process.}
  \label{fig:1.2_Merge_Datasets}
\end{figure}

Following dataset generation, the Data Merger component aggregates selected app categories into a consolidated dataset, which is directly split into a test and a train dataset. Users can configure the following parameters (Figure \ref{fig:1.2_Merge_Datasets}):

\begin{itemize}

\item \textbf{Dataset Selection:} Users choose a dataset generated by the Data Selector.

\item \textbf{Category Merging:} Users specify which categories should be merged for analysis.

\item \textbf{Test-Train Split:} Users define the ratio for the test and training split, generating subsets of the dataset.

\end{itemize}

The resulting datasets are stored as a compressed zip file in MongoDB. These datasets are structured to accommodate subsequent AI model training and evaluation.

\subsection{Data Preprocessor}

\begin{figure}[h]
  \centering
  \includegraphics[width=0.55\linewidth]{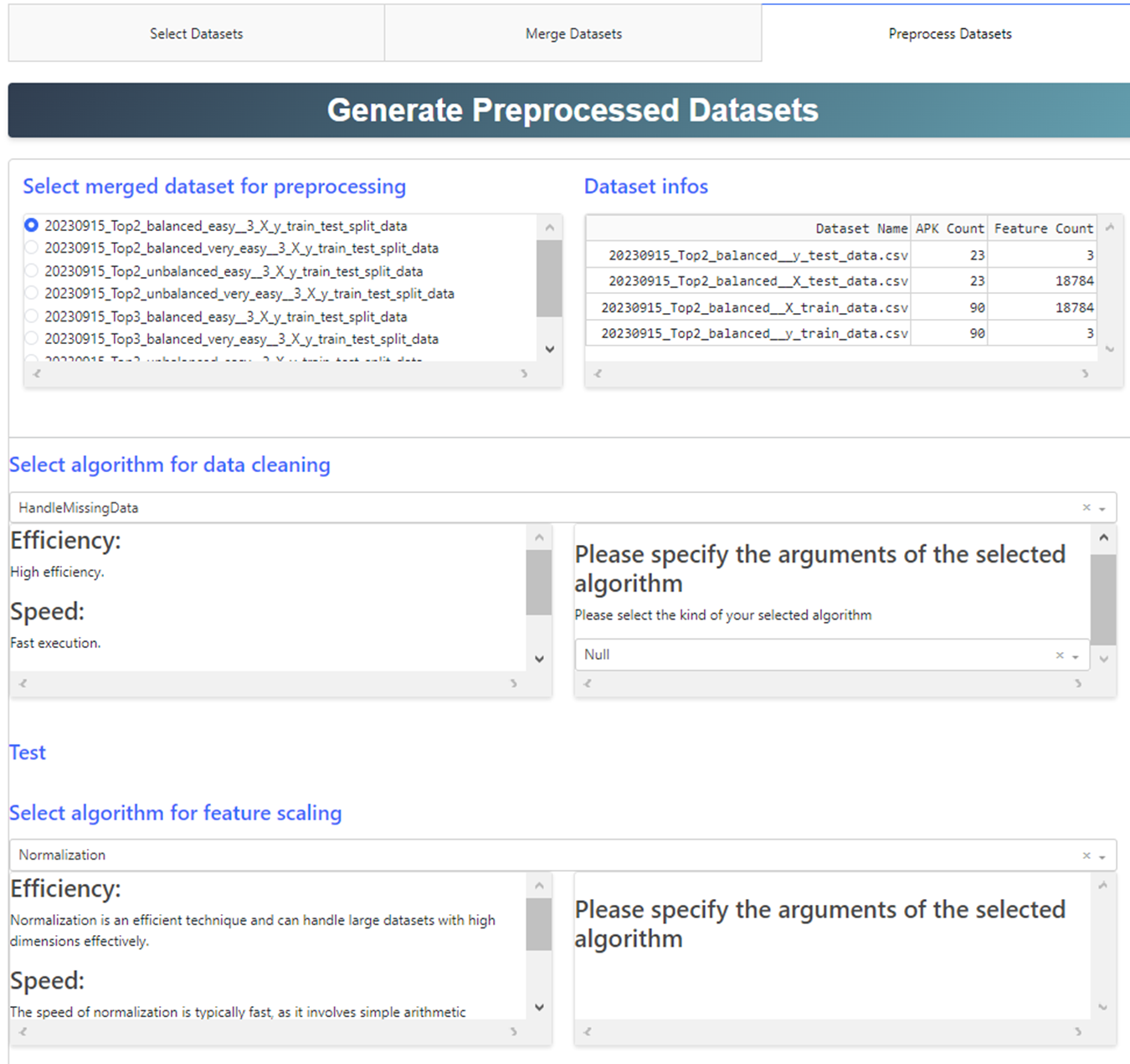}
  \caption{User Interface to Manage the Dataset Preprocessing Process.}
  \label{fig:1.3_Select_Datasets}
\end{figure}

The data preprocessor component is pivotal for refining the dataset. Users can select a merged dataset from the previous stage and apply a range of preprocessing algorithms to it (Figure \ref{fig:1.3_Select_Datasets}). From data cleaning and scaling to encoding, transformation, and feature selection, this stage empowers users to tailor data to their analysis needs. The algorithms can be configured with the following options:

\begin{itemize}

\item \textbf{Algorithm Selection:} Users choose from a set of available preprocessing algorithms, each designed to address specific data manipulation needs. To help inexperienced users, the system provides a short description of each algorithm.

\item \textbf{Hyperparameters:} Users can customize hyperparameters associated with selected algorithms, adapting preprocessing techniques to the dataset's nuances.

\item \textbf{Feature Sensitivity:} Certain algorithms are feature-sensitive, while others operate across all features.

\item \textbf{Dataset Naming:} A unique name for the dataset is provided by the user, ensuring simple identification and future reference.

\end{itemize}

The selected preprocessing algorithms are executed on the dataset, resulting in a refined dataset that has undergone data cleaning, scaling, encoding, transformation, extraction, selection, combination, and other relevant preprocessing steps. The resulting dataset is compressed and stored. Furthermore, the user inputs and preprocessing choices are stored for reproducibility.
The data preprocessing pipeline, by integrating the data selector, data ,erger, and data preprocessor components, facilitates the preparation of structured and refined datasets primed for AI model training.

%Nameing is missed in the preview stages

\section{AI Model Pipeline (Stage 3)}
\label{sec:ai_model_pipeline}

The AI Model Pipeline represents the culmination of the ALPACA system and includes training, evaluation and prediction. The choice of AI model depends on the underlying use case. However, the models are designed so that they can be used across domains and are not linked to the existence of specific features. According to its dockerized, Celery-based design, multiple AI model workers can run simultaneously - depending on the available hardware resources. This chapter delves into the intricacies of the AI model pipeline and explains the role of training, evaluation, and prediction in generating insights and predictions from preprocessed and refined datasets. It also covers the methods, algorithms, and evaluation strategies used to train AI models, analyze their performance, and provide users with actionable insights.

\begin{figure}[h]
    \centering
    \includegraphics[width=0.7\linewidth]{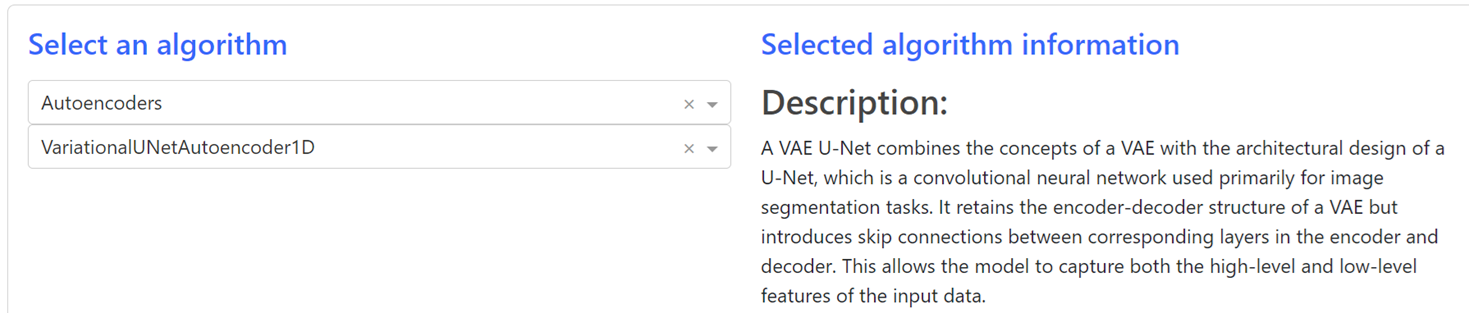}
    \caption{How to Select an Algorithm.}
    \label{fig:user_helpert}
\end{figure}

\subsection{Model Training}

\begin{figure}[h]
  \centering
  \begin{minipage}{0.4\linewidth}
    \centering
    \includegraphics[width=0.9\linewidth]{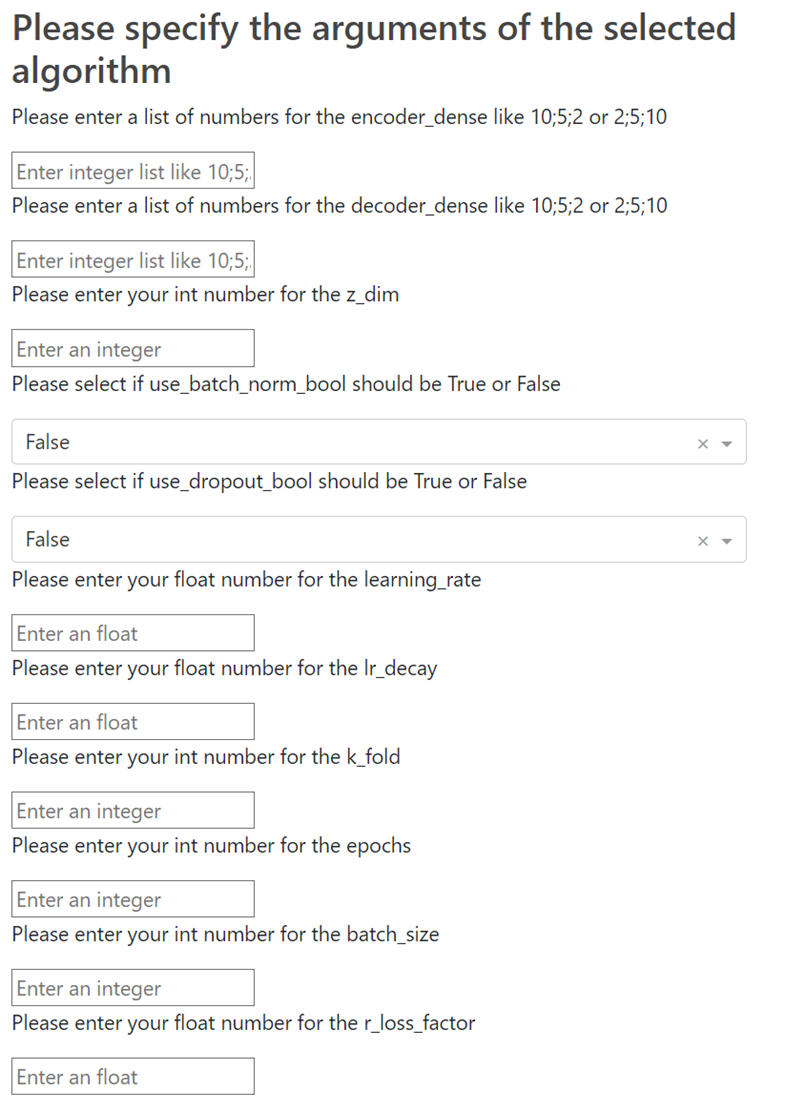}
    \caption{Alpacas' AI Algorithm Configuration.}
    \label{fig:User_input}
  \end{minipage}
  \hfill
  \begin{minipage}{0.4\linewidth}
    \centering
    \includegraphics[width=0.9\linewidth]{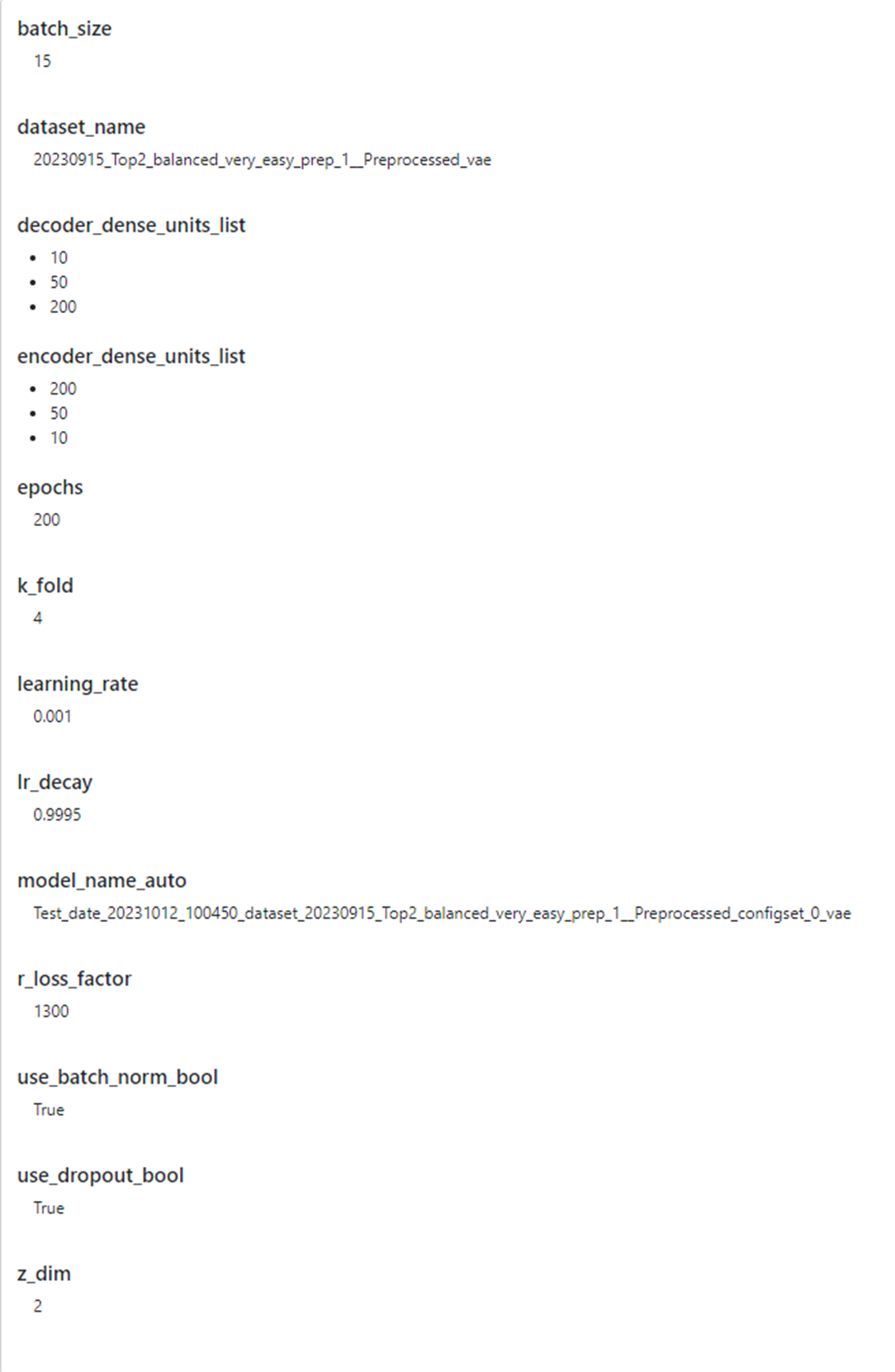}
    \caption{Stored User Input.}
    \label{fig:User_input_tracking}
  \end{minipage}
\end{figure}

The model training component serves as the core of the AI model pipeline, orchestrating the training of AI models on the preprocessed datasets. Users can customize their model training by parametrizing the following:

\begin{itemize}

\item \textbf{Dataset Selection:} Users select one of the preprocessed datasets created in the previous phase, which they would then like to use for training.

\item \textbf{App Categories:} Depending on the analysis goals, users select the category in which the AI model will be trained.

\item \textbf{AI Algorithm Selection:} Users can choose from a range of AI algorithms. To do this, they first select a class of AI algorithms (e.g., autoencoder). They then decides on a specific algorithm from this class (e.g., VAE). Each of these algorithms offers different strengths and weaknesses for learning patterns and representations within the data. Users can experiment to see which algorithm produces the best results for specific use cases. To help inexperienced users, the system provides a short description of each algorithm (Figure \ref{fig:user_helpert}).

\item \textbf{Hyperparameter Configuration:} Users specify the hyperparameters required for the chosen algorithm and adapt the training process to the requirements and characteristics of their dataset (Figure \ref{fig:User_input}).
This includes, for example, parameters such as batch size, layer architecture, and learning rate, which significantly influence the behavior of the model during training. This allows users to experiment with the effects these parameters have.

\item \textbf{Model Naming:} A unique name for the trained model is provided by the user, ensuring easy identification and future reference.

\item \textbf{Model Training:} The training process can be initiated multiple times by the user to optimize the model's parameters for the chosen dataset. Training progress and outcomes are logged and monitored for transparency, making it easy to experiment with different datasets, algorithms, and hyperparameters.

\end{itemize}

The model training component leverages the specified dataset and algorithm to train the AI model. After training, the model will be evaluated and can then be used for predictions. The resulting model and all user inputs are saved for reproducibility. An example of the configuration of an algorithm and the corresponding stored input data can be seen in Figure \ref{fig:User_input_tracking}.

\subsection{Model Evaluation}

After model training, AI models are automatically subjected to rigorous evaluation to assess their performance and gain insights from the analysis. Some of the evaluation criteria correlate more strongly with the task to be solved than others. These evaluation components must be replaced, depending on the application. The assessment covers various aspects:

\begin{itemize}

\item \textbf{Classical Evaluation Metrics:} The model is evaluated with some classically used metrics like true positives (TP), true negatives (TN), false positives (FP), and false negatives (FN). Furthermore, some derived metrics are calculated, like accuracy, precision, recall, specificity, and F1-Score, as well as true and false positive and true and false negative rates.

\item \textbf{Model Prediction Analysis:} The AI model's predictions are analyzed in terms of their accuracy and reliability. Predicted categories are compared with real labels to determine the model's effectiveness.

\item \textbf{Reconstruction and Latent Representation:} The trained model's ability to reconstruct features from the test dataset is assessed, offering insights into its ability to capture essential patterns. Additionally, the latent representation (hyperplane of the encoder output) of test and training data is visualized using dimensionality reduction techniques like t-SNE or PCA.

\item \textbf{Cluster Analysis:} A k-means clustering algorithm is applied to the latent representations of test and training data and provides insights into the groupability of the resulting data points.

\end{itemize}

\subsection{Prediction and Visualization}

\begin{figure}[ht]
  \centering
  \begin{minipage}{0.49\linewidth}
    \centering
    \includegraphics[width=\linewidth, height=0.75\linewidth]{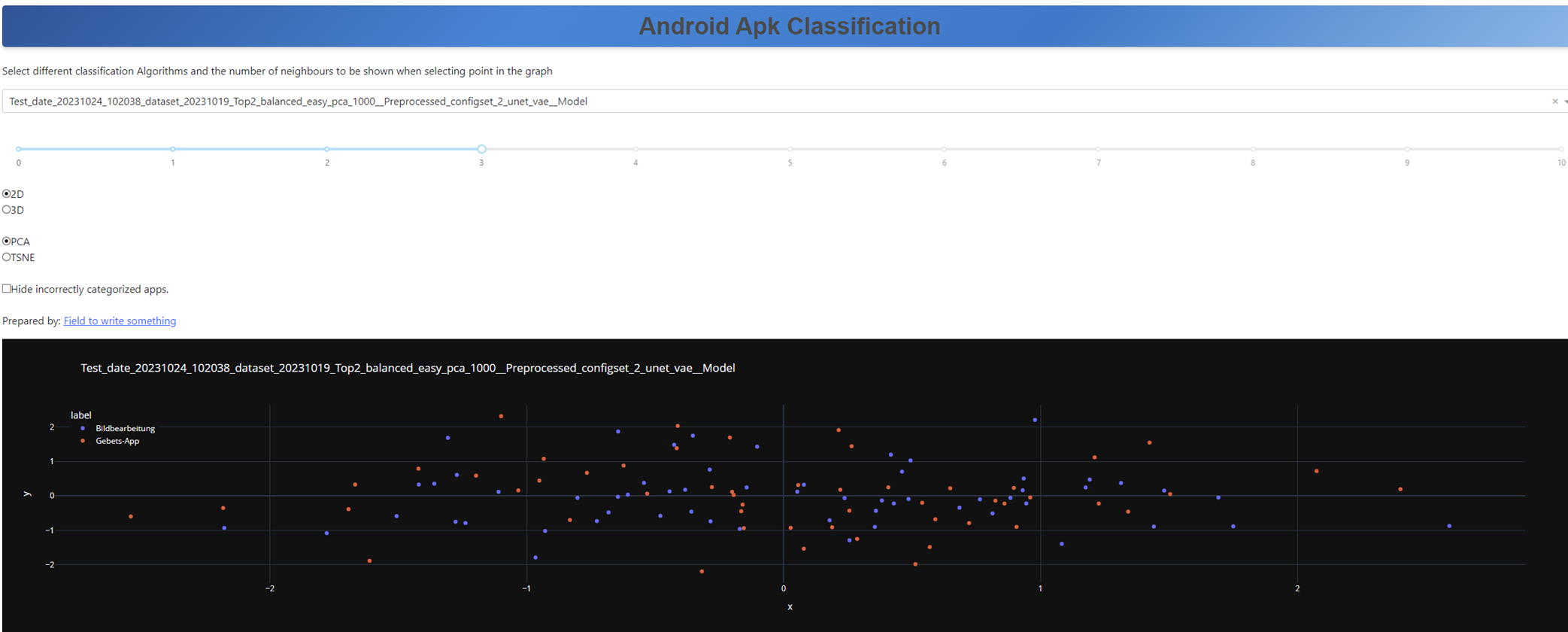}
    \caption{ALPACA Clustering Prediction View.}
    \label{fig:Predictor_large}
  \end{minipage}
  \hfill
  \begin{minipage}{0.49\linewidth}
    \centering
    \includegraphics[width=\linewidth, height=0.75\linewidth]{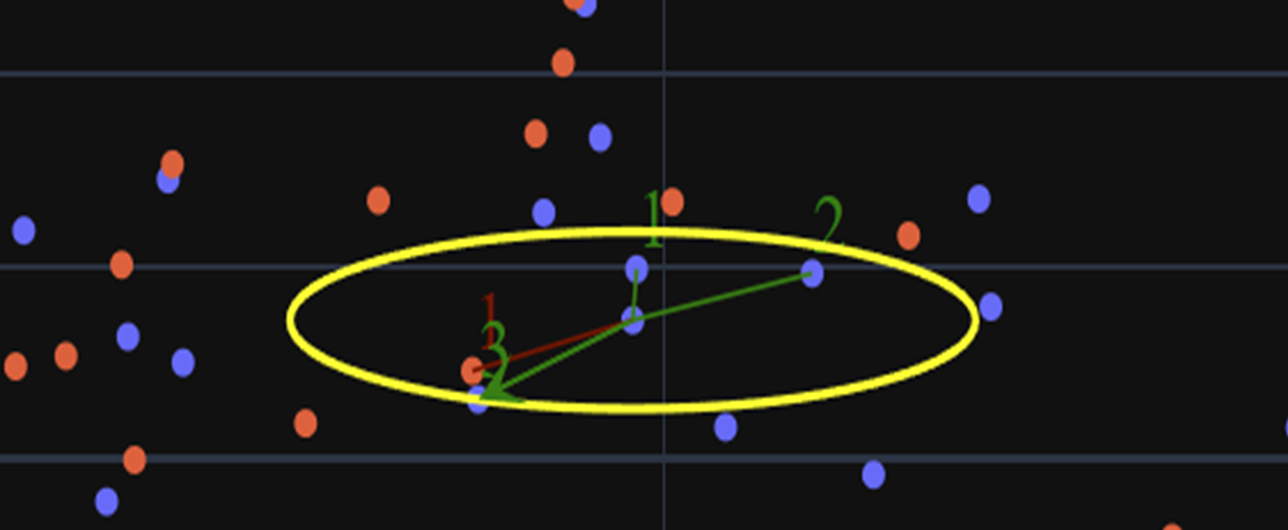}
    \caption{ALPACA Detail Clustering Prediction View.}
    \label{fig:Predictor_detail}
  \end{minipage}
\end{figure}

The ALPACA system provides a user-friendly frontend that empowers users to interact with trained AI models and gain insights\footnote{The user interface can be seen in Figure \ref{fig:Predictor_large}. To see how the prediction is shown after clicking an item a detailed picture can be seen in Figure \ref{fig:Predictor_detail}.}. The visualization of the model's output correlates with the model's use case and needs to be adjusted if necessary. The following parameters can be set for visualizing data:

\begin{itemize}

\item \textbf{Algorithm Selection:} Users choose the AI algorithm used for the prediction.

\item \textbf{Visualization Options:} Users can choose between 2D and 3D scatter plots, employing dimensionality reduction techniques for comprehensive visualization. Furthermore, users can adjust parameters like the number of nearest apps shown and toggle the display of incorrect predictions.

\item \textbf{Prediction:} Users can explore the AI model's predictions through interactive scatter plots. Apps are connected by arrows based on their predicted categories, aiding visualization of similarities and differences.

\item \textbf{Nearest Apps:} The system highlights the nearest apps in the scatter plot based on the app selected by the user and some user-desired properties, such as the number of neighbors to display, to help identify similar apps.

\item \textbf{Categorization Confidence:} The system uses arrow colors to indicate the categorization confidence of the predicted apps (green for the correct category, red for the incorrect category). The display of incorrectly categorized apps can be deactivated by the user.

\end{itemize}

\section{Legal and Ethical Considerations}
\label{sec:LegalandEthicalConsiderations}

The ALPACA system's capability raises important legal and ethical considerations regarding privacy regulations and copyright law. To ensure responsible and ethical handling of user and APK data, the ALPACA system integrates several processes and measures that ensure compliance with data protection regulations and copyright law.  These considerations will also influence the system's further development in the future and ensure responsible data use. For this reason, the system cannot be offered as open source.  However, the source code can be passed on upon request and after clarifying the legal framework.
\section{Conclusion}
\label{sec:Conclusion}

The introduction of artificial intelligence into everyday life and the resulting ever faster and more extensive development of AI systems lead to a constantly increasing need for systematic AI pipelines. However, to ensure the acceptance of AI, this disruptive technology must remain accessible, understandable and transparent for all user groups. Increasingly complex data and constant change also require adaptable and easy-to-maintain systems that can seamlessly integrate various phases such as data collection, preparation, model generation and evaluation. Collaborative, adaptive and user-friendly AI pipeline systems offer a solution. These systems consist of a complex system of interconnected data processing and analysis stages, each playing a unique role in transforming raw data into meaningful insights. This paper presented ALPACA (Adaptive Learning Pipeline for Advanced Comprehensive AI Analysis), which addresses these problems and presents a web-based AI pipeline that can support different user groups in the use of artificial intelligence. For this purpose, modules and artefacts can be exchanged between users. This creates an AI ecosystem that enables a broad range of people to participate in this topic.

ALPACA was designed as a universal, expandable AI pipeline framework that is aimed at various user groups from data scientists and AI experts to specialists in other fields as well as students and laypeople, fostering an ecosystem which connects them. The system design was chosen so that it can be expanded by experts without making any changes to the existing system. For this purpose, developers have several base classes and interfaces at their disposal. In addition, mechanisms for providing descriptions and explanations are implemented to support even inexperienced users. The graphical interfaces required for operation are automatically derived from the classes using a design based on reflections. On the one hand, this offers developers the opportunity to continuously improve the performance of the system and integrate new functions quickly and easily. On the other hand, this approach also ensures a consistent design of the back- and frontend and enables uniform use across all pipeline levels and all necessary tools. This makes both maintenance and usability easier. Users, on the other hand, can rely on the ever-growing range of AI tools and solve new challenges through the use of AI even without special knowledge. This means that previously unexplored areas can also be penetrated by AI.

Another important feature of ALPACA lies in its ability to use graphics processing units (GPUs) for data preprocessing and model training. This significantly increases computing speed, especially in complex network configurations, and ensures scalability and cost efficiency through cloud computing. The paper highlights the transformative potential of GPU-accelerated AI, making AI pipelines more efficient and accessible to a wider audience.

In addition, ALPACA offers solutions tailored to different user groups to meet their needs and ensure efficient use of the pipeline:

\begin{itemize}

\item \textbf{Data Scientists and AI Experts:} ALPACA optimizes workflows and enables efficient experimentation with different algorithms and hyperparameters. Securing intermediate results enables comprehensive collaboration, which increases knowledge exchange and productivity.

\item \textbf{Specialists in Other Areas:} Subject-specific experts leverage ALPACA's user-friendly interfaces and automated workflows, bridging the gap between their expertise and advanced AI techniques. This democratization of AI enables subject-matter experts to harness the power of AI in their respective areas of expertise.

\item \textbf{Students and Laypeople:} ALPACA serves as an educational platform and can be used for interactive tutorials, example projects and guided exercises. Exploratory learning allows people without specific AI knowledge to experiment with models and overcome real-world challenges. This will promote a culture of continuous learning and increase public acceptance of AI.

\end{itemize}

In addition, this article presents a demo scenario from the area of Android similarity detection, which shows the versatility of ALPACA. This practical application demonstrates ALPACA's ability to handle complex AI tasks while remaining accessible and understandable to a wide range of users. Additionally, the demo scenario shows how a crowdsourcing approach can solve the often complex problem of data categorization by harnessing the power of the crowd.

In summary, ALPACA represents a significant advance in the development of AI pipelines, offering a user-centric approach that embraces inclusivity, ease of use and collaboration. By addressing the challenges of complexity and accessibility, ALPACA is paving the way for a future in which AI becomes an integral part of various fields, thereby democratizing access to the transformative power of artificial intelligence. In addition, ALPACA's overall architectural design based on modularity, scalability and user-centered design proves to be a powerful and adaptable tool for comprehensive AI analysis. By seamlessly integrating data management, user interaction and ethical considerations, the ALPACA system allows users to participate in the analysis process while maintaining standards of transparency, privacy and reproducibility. Additionally, the paper highlights the critical role of user experience and usability in building trust in AI systems and highlights the need for domain-agnostic AI pipelines that can support and connect users across different knowledge levels. ALPACA achieves this through the harmonious integration of complex technical components and the convergent design of a behaviour-sensitive, visual user interface. This facilitates access and opens the door to an ecosystem where different user groups can experiment and contribute their expertise, as well as train, analyze, evaluate and share models.
\section{Future Work}
\label{sec:FuturWork}

The black-box nature of AI models raises concerns about their transparency and interpretability. To counteract this, the inclusion of \textbf{explainable AI} techniques in future a future version of ALPACA is essential. The integration of this technology would illuminate the decision-making processes of AI models in detail and ensure that decisions are understandable. In addition, this approach would further promote trust in such systems which is a main goal of ALPACA. The introduction of AI into critical areas such as healthcare, finance, and the legal sector will only be made possible by a systems which supports explainable A.

Incorporating concepts such as \textbf{Federated Learning} and \textbf{Continuous Learning} would further expand ALPACA's capabilities. Federated Learning, a decentralized approach to model training, allows devices or nodes to learn together while keeping the data localized, which would address privacy and security concerns. Continuous learning, on the other hand, would enable the models to adapt to changing circumstances. The future integration of these concepts into ALPACA will ensure a dynamic and adaptive AI system.

Continuing to \textbf{improve user interfaces} and continually \textbf{embedding robust ethical frameworks} based on new policies and laws are critical to ensure ALPACAs' usabillity and legality. These efforts will continue to make sure user privacy and ethical data use, creating a user-friendly and trustworthy environment in the long term.

As the Android ecosystem evolves, the effectiveness of ALPACA in terms of the demo scenario depends on the Androguard framework used for app analysis. Such an outdated \textbf{tools landscape} is subject to certain restrictions due to technical progress. In the case of AndroGuard, the last stable version was released on February 18, 2019 \cite{AndroGuard.Versions.GitHub.08.11.2023b}, in the age of Android Oreo (version 9, API level 28). However, for good results, the basic data of the model training must be current, precise, and correct. It is therefore crucial that the tools used for data collection are adapted to technical changes and advances. Androguard should either be updated or replaced for future projects.

In summary, these improvements would make ALPACA a more transparent, adaptable, and user-friendly AI analysis tool. Consideration and integration of these areas would ensure the relevance and impact of ALPACA in various user scenarios and critical applications in the future.
\section{Acknowledgement}
We thank our colleagues at the DAI-Lab of the TU Berlin for their helpful feedback and support. In particular, we would like to thank Karsten Bsufka for his valuable contributions to our research.

This work is supported in part by the German Federal Ministry of Education and Research (BMBF) under grant number 16SV8518.

% Use the style of numbering in square brackets.
% If nothing is used, default style will be taken.
%\begin{enumerate}[a)]
%\item 
%\item 
%\item 
%\end{enumerate}  

%\begin{itemize}
%\item 
%\item 
%\item 
%\end{itemize}  

%\begin{description}
%\item[]
%\item[] 
%\item[] 
%\end{description}  

% Figure
%\begin{figure}[<options>]
%	\centering
%		\includegraphics[<options>]{}
%	  \caption{}\label{fig1}
%\end{figure}

%\begin{table}[<options>]
%\caption{}\label{tbl1}
%\begin{tabular*}{\tblwidth}{@{}LL@{}}
%\toprule
%  &  \\ % Table header row
%\midrule
% & \\
% & \\
% & \\
% & \\
%\bottomrule
%\end{tabular*}
%\end{table}

% Uncomment and use as the case may be
%\begin{theorem} 
%\end{theorem}

% Uncomment and use as the case may be
%\begin{lemma} 
%\end{lemma}


%\section{}\label{}


\bibliographystyle{cas-model2-names}

\bibliography{sources}

\bio{}

\endbio

\endbio

\end{document}